\documentclass[english,aps,prb,superscriptaddress,reprint,longbibliograhy]{revtex4-2}
\usepackage[T1]{fontenc}
\usepackage[latin9]{inputenc}
\usepackage{babel}
\usepackage{graphicx}
\usepackage{bm}
\usepackage{amsmath}
\usepackage{amssymb}
\usepackage{amsfonts}
\usepackage{dcolumn}
\usepackage{bbold}
\usepackage{tikz}
\usepackage{xfp}
\usepackage{verbatim}

\usepackage[margin=15mm]{geometry}
\colorlet{linkequation}{blue}
\usepackage[colorlinks=true,linkcolor=cyan,citecolor=red,urlcolor=cyan]{hyperref}
\usetikzlibrary{shadings}

\usepackage{color}

\definecolor{goodred}{RGB}{183,15,58}
\definecolor{goodblue}{RGB}{93,128,180}
\definecolor{goodgreen}{RGB}{95,118,33}

\usepackage{color}

\begin{document}

\begin{abstract}
The dynamics of magnetic moments consists of a precession around the magnetic field direction and a relaxation towards the field to minimize the energy. While the magnetic moment and the angular momentum are conventionally assumed to be parallel to each other, at ultrafast time scales their directions become separated due to inertial effects. The inertial dynamics gives rise to additional high-frequency modes in the excitation spectrum of magnetic materials. 
{ Here, we}  
review the recent theoretical and 
experimental advances in this emerging topic and discuss the open challenges and opportunities 
in the detection and the potential applications of inertial spin dynamics.     
\\



\end{abstract}

\title{Inertial effects in ultrafast spin dynamics}

\author{Ritwik Mondal}
\email[]{ritwik@iitism.ac.in}

\affiliation{Department of Physics, Indian Institute of Technology (ISM) Dhanbad, IN-826004, Dhanbad, India}


\author{Levente R\'ozsa}
 
\altaffiliation{Present address: Department of Theoretical Solid State Physics, Institute of Solid State Physics and Optics, Wigner Research Centre for Physics, H-1525 Budapest, Hungary}
\altaffiliation{Department of Theoretical Physics, Budapest University of Technology and Economics, H-1111 Budapest, Hungary}
\affiliation{Department of Physics, University of Konstanz, DE-78457 Konstanz, Germany}

\author{Michael Farle}
\affiliation{Faculty of Physics, University of Duisburg-Essen, DE-47048 Duisburg, Germany}


\author{Peter M. Oppeneer}
\affiliation{Department of Physics and Astronomy, Uppsala University, Box 516, SE-75120 Uppsala, Sweden }

\author{Ulrich Nowak}
\affiliation{Department of Physics, University of Konstanz, DE-78457 Konstanz, Germany} 

\author{Mikhail Cherkasskii}
\email[]{macherkasskii@hotmail.com}
\altaffiliation{Present address: Institute for Theoretical Solid State Physics, RWTH Aachen University, DE-52074 Aachen, Germany}
\affiliation{Faculty of Physics, University of Duisburg-Essen, DE-47048 Duisburg, Germany}

\maketitle
\date{\today}

\section{Introduction}
The increasing challenge of processing and storing a 
rapidly growing amount of digital information requires novel technological solutions operating at smaller length scales and at increased speed, yet in a more energy-efficient manner. 
While current magnetic devices enable data storage on short length scales with a low energy consumption, reading and rewriting the bits using magnetic field pulses~\cite{SIEGMANN1995L8} is not possible below the nanosecond time scale. 


To manipulate the spins on shorter time scales, electrical currents and ultrafast optical laser pulses have been employed. These methods enable ultrafast demagnetization within femtoseconds~\cite{Bigot1996} and magnetization switching within picoseconds in a broad variety of magnetic materials~\cite{Stanciu2007,Radu2011,ostler12,LeGuyader2015,Mangin2014,KOPLAK2017}. Many aspects of ultrafast demagnetization and switching can be successfully described either phenomenologically~\cite{Vahaplar2009,Mentink2012}, or microscopically based on the Landau--Lifshitz--Gilbert (LLG) equation~\cite{landau35,Gilbert2004} in its stochastic form~\cite{brown1963micromagnetics,Barker2013,Wienholdt2013}. While the latter approach is widely applied to modelling magnetization dynamics in the presence of thermal fluctuations, it relies on the crucial assumption that the spin degrees of freedom are coupled to a heat bath responsible for the dissipation as well as the thermal noise, while details of the considerably faster electronic and lattice degrees of freedom constituting the heat bath are neglected~\cite{brown1963micromagnetics,Antropov1996}. Recent derivations of the LLG equation based on a relativistic theory~\cite{Mondal2017Nutation,Mondal2018JPCM} have proven that this approximation is no longer justified if the spin directions significantly vary over the course of femtoseconds. 

At ultrashort time scales, the LLG equation has to be corrected by accounting for the fact that the magnetization direction can no longer instantaneously follow the angular momentum. This delay can be described by appending an inertial term including the second time derivative of the magnetization to the LLG equation~\cite{Suhl1998,wegrowe2000thermokinetic,Ciornei2011,Wegrowe2012,giordano2020derivation,Jingwen2022}. This phenomenological consideration is supported by various derivations of the inertial term based on microscopic relativistic quantum theories~\cite{Fahnle2011,Bhattacharjee2012,Mondal2017Nutation,Mondal2018JPCM}. There are numerous theoretical predictions on how the signatures of inertial dynamics can be detected, but experimental observations are limited so far. Most likely this can be attributed to the fact that conventional magnetic measurements focus on the low-frequency regime, typically on the GHz range in ferromagnets, where the inertia plays little role and its effects may alternatively be explained based on the conventional LLG equation. However, the magnetic moments not only experience precession around the effective field in the presence of the inertial term, but they also perform a high-frequency nutation around the angular momentum, see Fig.\,\ref{fig:Fig1}. Hence, the nutation gives rise to an additional peak in the ferromagnetic resonance spectrum in the high-frequency regime~\cite{Olive2012}. 
This resonance is typically found in the THz range in contrast to the conventional precession resonance at GHz frequencies. The most convincing experimental signatures of inertial dynamics to date are based on the observation of this high-frequency response in NiFe, CoFeB~\cite{neeraj2019experimental} and Co~\cite{unikandanunni2021inertial} films. {{Propagating nutational spin waves have also been predicted to possess frequencies in the THz regime~\cite{Kikuchi}, but have not been observed experimentally so far.}} 


In this review, we first describe the inertial LLG equation by motivating the precession, damping and inertial terms. We discuss the consequences of inertial dynamics on resonance spectra, on the spin-wave dispersion and on switching processes not only in ferromagnets, but also in antiferromagnets and ferrimagnets. We also outline the challenges and opportunities concerning the experimental observation of inertial spin dynamics, paving the way towards a microscopic understanding and possible technological applications of the evolution of magnetic moments on ultrafast time scales.


\section{Magnetization dynamics}

Here, we summarize the main {{points}} of LLG dynamics, and point out in which aspects it has to be modified at ultrashort time scales, culminating in the formulation of the inertial LLG equation.

\subsection{Precession and damping dynamics}
{When a 
magnetic moment {$\boldsymbol{ M}_{0}$} is placed in an external magnetic field $\boldsymbol{ B}$, the corresponding energy is {$\mathcal{H}=-\boldsymbol{ M}_{0}\cdot \boldsymbol{ B}$}}.
The energy is minimized when the direction of the magnetic moment is parallel to the direction of the magnetic field. 
In classical electrodynamics, the magnetic moment is represented by a charged particle moving along a closed curve, establishing a relation between its angular momentum {$\boldsymbol{L}_{0}$} and magnetic moment {$\boldsymbol{M}_{0}$} via the relation {$\boldsymbol{M}_{0} = \gamma \boldsymbol{L}_{0}$}, where $\gamma$ is the gyromagnetic ratio. 
The rate of change of angular momentum is equal to the torque, leading to the precessional motion of the magnetic moment~\cite{blundell01}
\begin{align}
    \dot{\boldsymbol{M}}_{0} = -\gamma \boldsymbol{M}_{0}\times \boldsymbol{B}\,,
    \label{Eq1}
\end{align}
for electrons with a negative charge $-e$ and mass $m$. 
Note that the magnitude of the magnetic moment $M_{0}$ 
remains constant. 
An identical equation of motion may be derived by treating the moment quantum-mechanically. The only difference is in the value of the gyromagnetic ratio $\gamma=ge/\left(2m\right)$, where the gyromagnetic factor is $g=1$ for classical particles and is close to $g\approx 2$ for electrons in a solid where the quantum-mechanical spin angular momentum is the dominant contribution to the magnetic moment. The value $\gamma=1.76\cdot 10^{11}$\,T$^{-1}$s$^{-1}$ for $g=2$ sets the 
characteristic frequencies of magnetic moment dynamics in the gigahertz range for typically achievable magnetic field values of a few Tesla.

It is known that the magnetic moment of ferromagnets does not only precess around the field, but also minimizes its energy by becoming parallel to it within microscopic time scales. To incorporate this experimental fact, adding a phenomenological damping term to the equation of motion was suggested by Landau and Lifshitz~\cite{landau35}. An alternative formulation of the damped equation of motion was proposed by Gilbert~\cite{Gilbert2004}
, setting an upper bound on the damping coefficient in the Landau--Lifshitz formalism to better accommodate experimental observations. 
For an ensemble of interacting magnetic moments, the magnetization dynamics can be described by the Landau--Lifshitz--Gilbert (LLG) equation of motion 
\cite{landau35,Gilbert2004},
\begin{align}
    \dot{\boldsymbol{M}}_{ i}(t) = - \gamma_{ i} \boldsymbol{ M}_{ i} \times \boldsymbol{ B}^{\rm eff}_{ i} + \frac{\alpha_{ i}}{M_{ \textrm{0},i}}  \boldsymbol{ M}_{ i} \times \dot{\boldsymbol{ M}}_{ i} \,,  \label{Eq2}
\end{align}
where ${ i}$ stands for the indices of the magnetic moments and $M_{\textrm{0},i}$ are the magnitudes of the moments, which are still conserved during the time evolution. 
$\alpha_{ i}$ are the Gilbert damping parameters that phenomenologically describe the energy dissipation in terms of the coupling of the magnetic moments to the considerably faster degrees of freedom. 
The typical frequency scale of the dissipation is given by $\alpha_{i}\gamma_{i}/\left(1+\alpha_{i}^{2}\right)B$, which is usually much slower than the precession dynamics for common values of $\alpha_{i}\sim 10^{-5}-10^{-2}$. Equation~\eqref{Eq2} can be readily rewritten for a continuous magnetization field $\boldsymbol{M}(\boldsymbol{r})$ with saturation value $M_{ \textrm{S}}$, as originally proposed by Landau and Lifshitz~\cite{landau35}. The effective field $\boldsymbol{ B}^{\rm eff}_{ i}$ can be calculated from the Hamiltonian $\mathcal{H}$ of a magnetic system following the definition $\boldsymbol{ B}^{\rm eff}_{ i} = -\partial\mathcal{H}/\partial \boldsymbol{ M}_{ i}$ in the discrete case, replaced by the free energy $\mathcal{F}$ and $\boldsymbol{ B}^{\rm eff}_{ i} = -\delta\mathcal{F}/\delta \boldsymbol{ M}$ in the continuum limit. The Hamiltonian contains interactions of the magnetic moments with the external field through the Zeeman term, with the atomic lattice through magnetocrystalline anisotropy terms, and with each other in the form of dipolar and exchange interactions.

Though the original LLG equation is based on a phenomenological description~\cite{landau35,Gilbert2004}, several theories on the microscopic origins of the Gilbert damping have been put forward. In particular, the Gilbert damping has been proposed to originate from the breathing Fermi surface model \cite{kambersky70}, the torque-torque correlation model \cite{kambersky76,kambersky07,Thonig2018}, scattering theory formalism \cite{Brataas2008}, linear-response theory \cite{EbertPRL2011}, and relativistic Dirac theory \cite{Mondal2016,Mondal2018PRB}. The damping coefficient has also been generalized to a tensor~\cite{Brataas2008,Thonig_2014,Mondal2016,Thonig2018,Mondal2018PRB}, which is responsible for anisotropic damping observed in experiments~\cite{platow1998correlations, Farle2013}. Since the magnetic moment primarily stems from the spin angular momentum while the damping describes coupling to the lattice degrees of freedom, a common point of these microscopic theories is that the damping originates from the spin--orbit coupling. 

{The LLG equation~\eqref{Eq2}, which describes the dynamics of the mean value of the magnetization, can be augmented to incorporate the effects of thermal fluctuations. Brown proposed~\cite{Brown_1963,coffey2004langevin} 
that a thermal noise term should be added to the effective field such that $\boldsymbol{ B}^{\rm eff}_{ i} = -\frac{\partial \mathcal{H}}{\partial{ M}_{ i}} + \boldsymbol{\zeta}_{ i}(t)$, turning it into a stochastic differential equation.} This approach was generalized to interacting spin systems later~\cite{Lyberatos1993,CHUBYKALO2003}. Assuming that the system follows the Boltzmann distribution in thermal equilibrium, this noise term has the following properties: 
\begin{align}
    \langle \zeta_{i\eta}(t)\rangle & = 0\label{Eq3}\\
    \langle \zeta_{i\eta}(t) \zeta_{j\theta}(t^\prime) \rangle & = \delta_{ij}\delta_{\eta\theta} \delta(t-t^\prime) \frac{2\alpha_{i} k_{\textrm{B}} T }{\gamma_{i} M_{0,i}}
    \label{Eq4}
\end{align}
where $\eta$ and $\theta$ denote Cartesian components, 
$k_{\textrm{B}}$ is the Boltzmann constant and $T$ denotes the temperature of the system. This corresponds to white noise with zero expectation value, which is uncorrelated in space, time and Cartesian components. Similarly to the damping term, the noise describes coupling to the faster 
electronic degrees of freedom, which can be considered to be uncorrelated at the time scale of the spin dynamics. Regarding the phononic degrees of freedom, the separation of time scales is less straightforward, and the microscopic description of the coupling between spins and phonons is a subject of current research~\cite{Tauchert2022}. The connection between dissipation and fluctuations is also expressed by the Einstein relation~\eqref{Eq4}. An alternative form of the stochastic LLG equation was proposed by Kubo and Hashitsume~\cite{Kubo1970}, primarily differing in the scaling of the parameters from Brown's formulation, similarly to the Landau--Lifshitz and Gilbert forms of the LLG equation. If it is assumed that the heat bath consisting of phonons and electrons evolves at faster time scales than the spin system, including a white noise in the equation is justified. 
However, such a separation and averaging out becomes invalid for femtosecond magnetization dynamics because the electron relaxation time in metals is on the order of 10 fs~\cite{Stohr2006}. Using a stochastic field with a coloured noise may be more accurate in such cases~\cite{Atxitia2009}.        




\subsection{Inertial dynamics}

As emphasized above, both the dissipation and the thermal noise term in the stochastic LLG equation were introduced under the assumption that the relatively slow motion of the magnetic moments
is only influenced by an average of the other degrees of freedom. At shorter time scales, additional effects have to be included in the equation of motion. First, describing the evolution of the magnetic moments on time intervals comparable to the time between electron and phonon scattering events requires going beyond the instantaneous values of the magnetic moments in the LLG equations by including memory effects~\cite{Suhl1998,Fahnle2011}. 
Second, it has been already pointed out by Gilbert~\cite{Gilbert2004} that although precession also exists in classical mechanics, the correspondence between the dynamics of a magnetic moment and a spinning top is incomplete since the former does not possess a physical inertial tensor when described by the LLG equation. Third, at these time scales the excitation energies of the magnetic moments become comparable to those of electronic excitations, requiring a common quantum treatment of the degrees of freedom.

\subsubsection{Classical theory}

\begin{figure}[tbh!]
    \centering
    \includegraphics[scale = 0.9]{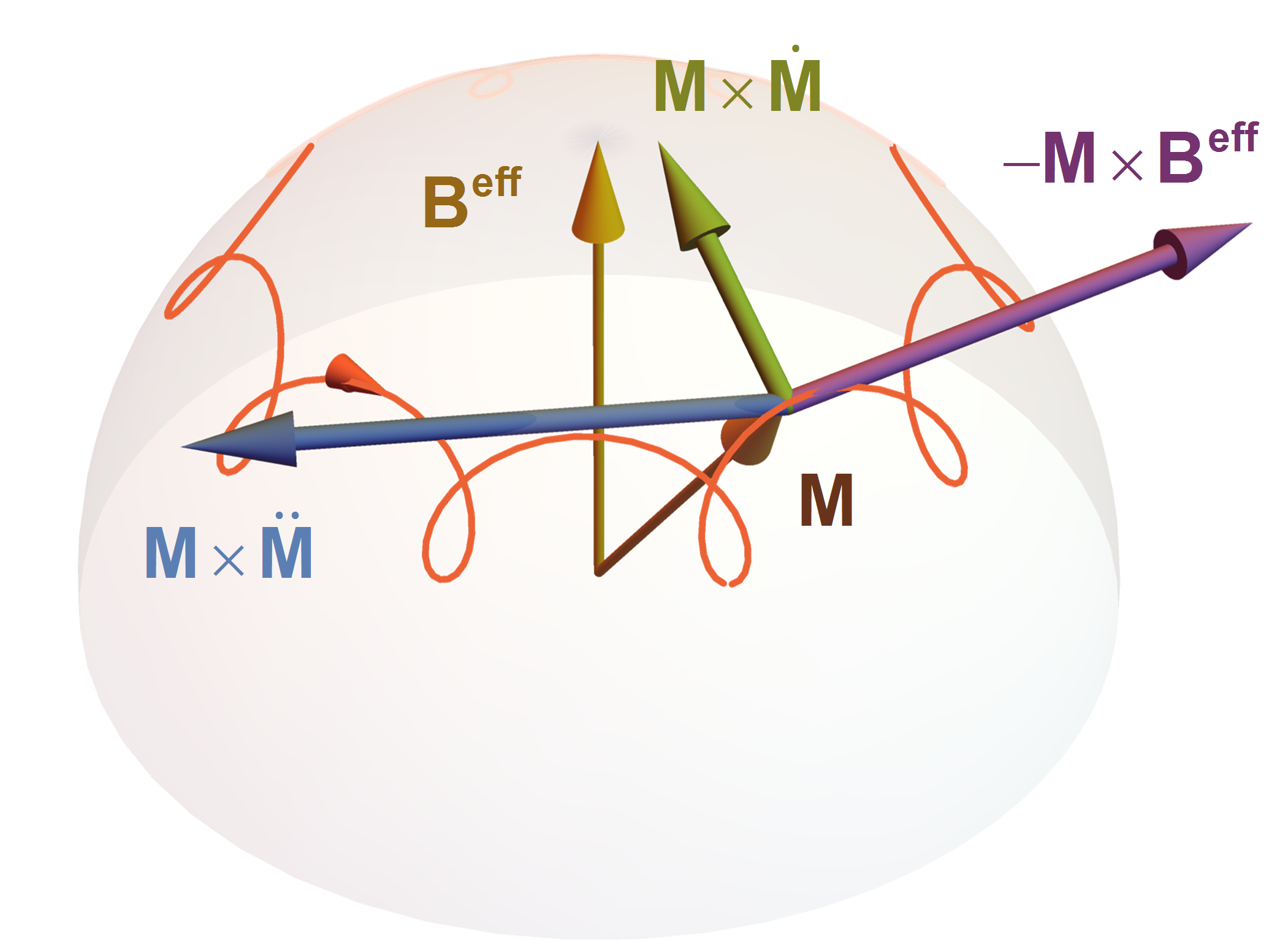}
    \caption{Schematic diagram of ILLG spin dynamics displaying the torques responsible for spin precession (purple), damping (green) and nutation (blue)
    ; from Ref.~\cite{Cherkasskii2022Anisotropy}.  
    }
    \label{fig:Fig1}
\end{figure}

While quantum electrodynamics provides an accurate description of the motion at high energy scales, here we discuss time scales ranging from 1~fs to 1~ps where a quasiclassical description remains valid. Both memory effects and the problem of the inertial tensor of magnetic moments may be treated by adding a second time derivative to Eq.~\eqref{Eq2}, resulting in the inertial LLG (ILLG) equation
\begin{equation} \label{Eq5}
\begin{split}
    \dot{\boldsymbol{ M}}_{ i}(t) &= - \gamma_{ i} \boldsymbol{ M}_{ i} \times \boldsymbol{ B}^{\rm eff}_{ i} + \frac{\alpha_{ i}}{M_{ 0,i}}  \boldsymbol{ M}_{ i} \times \dot{\boldsymbol{ M}}_{ i} \\
    &+ \frac{\eta_{ i}}{M_{ 0,i}}  \boldsymbol{ M}_{ i} \times \ddot{\boldsymbol{ M}}_{ i} \,,  
\end{split}
\end{equation}
Here, $\eta_{i}$ is the inertial relaxation time~\cite{neeraj2019experimental}, and the $\boldsymbol{ M}_{ i} \times \ddot{\boldsymbol{ M}}_{ i}$ form of the last term ensures the conservation of the length of the magnetic moments.

Memory effects can be fully treated by transforming the LLG equation into an integro-differential equation, as was derived in Ref.~\cite{Suhl1998} for spin-lattice and in Refs.~\cite{Fahnle2011,fahnle2013erratum} for spin-electron coupling. These types of equations are difficult to treat even numerically and require an expansion of the time integral, which leads to the damping and inertial terms containing first and second time derivatives, respectively. Third-order time derivatives were also included in Ref.~\cite{Suhl1998}, although it was emphasized that this form of the equation is not applicable at high frequencies. Indeed, higher-order derivatives are expected to lead to causality breaking, as is also known from the example of the Abraham--Lorentz force in electrodynamics. 

{{An alternative approach to derive Eq.~\eqref{Eq5} is based on the mechanical analogy with rigid-body motion, as is explained in detail in Refs.~\cite{Ciornei2011,Wegrowe2012}. The magnetic moment is pictured as a symmetric top, where $\boldsymbol{M}_{i}/M_{0,i}$ describes the direction of the axis of the top, which is now allowed to deviate from the direction of the angular momentum $\boldsymbol{ L}_{i}$. In the rotating frame where the axis of the top is fixed along the $\hat{\boldsymbol{z}}$ direction, the inertial tensor reads }}
\begin{align}
    \overline{\overline{I_{i}}} & = \begin{pmatrix}
    I_{i,1} & 0 & 0\\
    0 & I_{i,1} & 0\\
    0 & 0 & I_{i,3}
    \end{pmatrix}\, ,
\end{align}
and the connection between the angular momentum $\boldsymbol{ L}_{i} = (L_{i,1},L_{i,2},L_{i,3})$ and the angular velocity $\boldsymbol{ \Omega}_{i} = (\Omega_{i,1},\Omega_{i,2},\Omega_{i,3})$ is given by
$\boldsymbol{ L}_{i} = - \overline{\overline{I_{i}}} \boldsymbol{ \Omega_{i}}$, where the negative sign is introduced to follow the sign convention for $\gamma_{i}$ used here. The direction of $\boldsymbol{M}_{i}$ follows the time evolution
\begin{align}
    \dot{\boldsymbol{ M}_{i}} & = \boldsymbol{ \Omega}_{i} \times \boldsymbol{ M}_{i} \,,
\end{align}
by rotating with angular velocity $\boldsymbol{\Omega}_{i}$.
Taking a cross product on both sides with $\boldsymbol{ M}_{i} = M_{0,i}\hat{\boldsymbol{ z}}$, utilizing the double cross product $\boldsymbol{ M}_{i}\times \left(\boldsymbol{ \Omega}_{i}\times \boldsymbol{ M}_{i}\right) = \boldsymbol{ \Omega}_{i} M_{0,i}^2 - \boldsymbol{ M}_{i} \left(\boldsymbol{ \Omega}_{i} \cdot \boldsymbol{ M}_{i}\right) = \boldsymbol{ \Omega}_{i} M_{0,i}^2 - M_{0,i}^2\Omega_{i,3} \hat{\boldsymbol{ z}} $ and multiplying by $-\overline{\overline{I_{i}}}$, one obtains
\begin{align}
    \boldsymbol{ L}_{i} & = \frac{1}{\gamma_{i}}\boldsymbol{ M}_{i}-\frac{\eta_{i}}{\gamma_{i}M_{0,i}}\boldsymbol{ M}_{i}\times \dot{\boldsymbol{ M}}_{i}\, , \label{Angular_momentum}
\end{align}
where the notations $M_{0,i}/\gamma_{i}=-I_{i,3}\Omega_{i,3}$ and $\eta_{i}/\gamma_{i}=I_{i,1}/M_{0,i}$ were introduced. The time evolution of the angular momentum is governed by the precession and damping torques known from the LLG equation,
\begin{align}
    \dot{\boldsymbol{L}}_{ i}(t) = - \boldsymbol{ M}_{ i} \times \boldsymbol{ B}^{\rm eff}_{ i} + \frac{\alpha_{ i}}{\gamma_{i}M_{ 0,i}}  \boldsymbol{ M}_{ i} \times \dot{\boldsymbol{ M}}_{ i} \,.  \label{EqL}
\end{align}
Substituting Eq.~\eqref{Angular_momentum} into Eq.~\eqref{EqL} yields the ILLG equation \eqref{Eq5}.

In Gilbert's derivation of the LLG equation, $I_{i,1}$ was set to zero while $I_{i,3}$ was finite, which cannot occur for any mechanical rigid body~\cite{Gilbert2004}. For a finite $I_{i,1}$ or $\eta_{i}$, the angular momentum and the axis of the spinning top identified with the magnetic moment direction are no longer parallel to each other, and $\boldsymbol{M}_{i}$ performs a fast nutation around $\boldsymbol{L}_{i}$. {It is interesting to note that these fast and slow degrees of freedom are well separated \cite{Wegrowe2016JPCM}.} A schematic diagram of the ILLG equation is shown in Fig.~\ref{fig:Fig1}, displaying spin precession, relaxation and nutation. 

The ratio $\eta_{i}/\gamma_{i}$ stemming from the moment of inertia $I_{i,1}$ must necessarily be positive, which supports the interpretation of $\eta_{i}$ as an inertial relaxation time. 
This coefficient enables the introduction of the kinetic energy term
\begin{align}
\mathcal{T}=\sum_{i}\frac{\eta_{i}}{2\gamma_{i}M_{0,i}}\dot{\boldsymbol{M}}_{i}^{2}\, ,\label{eqT}
\end{align}
the lack of which was also pointed out by Gilbert for $I_{i,1}=0$. Taking the cross product of Eq.~\eqref{Eq5} with $\boldsymbol{M}_{i}$, then the scalar product with $\dot{\boldsymbol{M}}_{i}$ results in 
\begin{align}
\dot{\mathcal{T}}+\dot{\mathcal{H}}+\sum_{i}\frac{\alpha_{i}}{\gamma_{i}M_{0,i}}\dot{\boldsymbol{M}}_{i}^{2}=0\, ,\label{eqE}
\end{align}
describing the conservation of the total energy $\mathcal{T}+\mathcal{H}$ in the absence of damping~\cite{Mondal2020nutation}. The difference $\mathcal{T}-\mathcal{H}$ corresponds to the Lagrangian~\cite{Wegrowe2012}. 

Inertial magnetization dynamics of ferromagnetic nanoparticles including thermal excitations was investigated in Ref.~\cite{Titov2021Inertial}. It was found that adding the thermal noise term $\zeta_{i}$ with the moments given by Eqs.~\eqref{Eq3} and \eqref{Eq4} to the effective field can correctly account for the thermal fluctuations within the ILLG equation as well. The equilibrium Boltzmann distribution is defined by the sum of the kinetic and potential energies $\mathcal{T}+\mathcal{H}$ in this case, instead of only the potential energy for the stochastic LLG equation. The shorter time scales of inertial dynamics support the arguments in favour of replacing the white noise with a coloured noise~\cite{Atxitia2009}, which has not been considered in the ILLG formalism so far.

\subsubsection{Microscopic theory}

On a microscopic level, the ILLG equation has been derived from an extension of the breathing Fermi surface model~\cite{Fahnle2011,fahnle2013erratum}, from the torque-torque correlation model~\cite{Thonig2017}, as well as in atomistic~\cite{Bhattacharjee2012} and in Dirac relativistic quantum~\cite{Mondal2017Nutation,Mondal2018JPCM} frameworks. The latter approach is based on the derivation of a Pauli--Schr\"{o}dinger Hamilton operator from the Dirac equation,
\begin{align}
    \mathcal{H}_{\rm FW} & = \frac{\left(\boldsymbol{ p}-e\boldsymbol{ A}\right)^2}{2m} + V  - \frac{e\hbar}{2m}\, \boldsymbol{ \sigma}\cdot \boldsymbol{ B}\nonumber\\
    & + \mathcal{O}\left(\frac{1}{m^2c^2}\right) + \mathcal{O}\left(\frac{1}{m^3c^4}\right) + \dots 
    \label{FW_Hamiltonian}
\end{align}
by applying the Foldy-Wouthuysen transformation~\cite{foldy50}. The Zeeman term $e\hbar/\left(2m\right) \boldsymbol{ \sigma}\cdot \boldsymbol{ B}$ is responsible for the precession, where $\boldsymbol{ \sigma}$ denotes the vector of Pauli matrices. The spin-dependent part of the first-order relativistic correction term $\mathcal{O}\left(\frac{1}{m^2c^2}\right)$ results in the Gilbert damping, which contributes to the imaginary part of the magnetic susceptibility or the finite lifetime of excitations. The spin-dependent part of the second-order relativistic correction $\mathcal{O}\left(\frac{1}{m^3c^4}\right)$ includes higher-order spin--orbit coupling terms, and leads to intrinsic inertial dynamics~\cite{Mondal2017Nutation,Mondal2018JPCM} modifying the real part of the susceptibility. {
The intrinsic Gilbert damping parameter $\alpha_i$ and inertial relaxation time $\eta_i$ 
are generally considered to be constant in Eq.~\eqref{Eq5}. 
However, we emphasize that $\alpha_i$ and $\eta_i$ have to be time-dependent for pulsed, non-harmonic applied fields~\cite{Mondal2016,Mondal2017Nutation}, since the ILLG equation with the constant parameters may not capture the expected dynamics in the ultrafast regime~\cite{Hammar2017,Bajpai2019}. }

One of the pivotal questions of inertial spin dynamics is the time scales on which it is applicable, defined by the inertial relaxation time $\eta_{i}$. {The experimentally determined and theoretically predicted values of $\eta_{i}$ are summarized in Table~\ref{tab:eta}.} Although the phenomenological theory is not capable of calculating $\eta_{i}$, values of around 1-100~fs were proposed in Refs.~\cite{Ciornei2011,Wegrowe2012}. A value close to a single femtosecond was proposed in Ref. \cite{Bhattacharjee2012}, 
and deduced from ferromagnetic resonance measurements of the precession frequency in Ref.~\cite{Li2015}. First-principles calculations in Ref.\ \cite{Thonig2017} obtained smaller {absolute} values of $\eta_i \approx 10^{-3}$ fs{{, while in Ref.~\cite{Bouaziz2019} inertial relaxation times typically in the range of $10-100$~fs have been determined from ab initio simulations of the dynamical magnetic susceptibility}}. The detection of the resonant excitation of nutation by time-resolved magneto-optical measurements reported in Ref.~\cite{neeraj2019experimental} arrived at a value for $\eta_i$ on the order of $100$~fs in our convention. A recent measurement on Co films gave a value $\eta_i/\alpha_i \approx 750$~fs \cite{unikandanunni2021inertial} which also suggests that $\eta_i\approx 100$~fs. The large deviation between the values is remarkable since although the methods were different, almost all of the ab initio calculations and the experiments were performed for the $3d$ transition metal ferromagnets Fe, Co and Ni or their alloys. Even less reassuring is the fact that the $\eta_{i}/\gamma_{i}$ values were found to be negative in certain cases~\cite{Suhl1998,fahnle2013erratum,Li2015,Thonig2017,Bouaziz2019}. While a negative $\eta_{i}/\gamma_{i}$ may be substituted in the linear-response regime as was done in Ref.~\cite{Li2015}, the complete non-linear ILLG equation~\eqref{Eq5} is not meaningful for $\eta_{i}/\gamma_{i}<0$, since the magnetic moments could accelerate infinitely to decrease their kinetic energy in Eq.~\eqref{eqT}. A similar restriction is obtained for the damping $\alpha_{i}/\gamma_{i}>0$, otherwise the dissipation term would increase the energy over time in Eq.~\eqref{eqE}. Note that although Refs.~\cite{Ciornei2011,Kikuchi} use an opposite sign convention for the precessional term, the $\alpha_{i}/\gamma_{i}$ and $\eta_{i}/\gamma_{i}$ ratios are positive as required by energetic considerations. The estimated values of $\eta_{i}\approx 1-100$~fs are comparable to electron relaxation times in metals. Since the quasiclassical Boltzmann equation has proven successful in describing the non-equilibrium distribution of electrons and phonons on similar time scales, the ILLG equation can similarly be expected to correctly account for magnetic moment dynamics in this regime.

\begin{table}[tbh!]
\caption{\label{tab:eta}{{Comparison between the values of the inertial relaxation time $\eta_{i}$ obtained from various experimental and theoretical methods.}}}
\begin{ruledtabular}
\begin{tabular}{|>{\centering}p{3cm}|>{\centering}p{2cm}|>{\centering}p{2cm}|}
\hline 
Sample & $\eta_{i}\:\left(\text{fs}\right)$ & Ref.\tabularnewline
\hline 
\hline 
\multicolumn{3}{|c|}{Experiment}\tabularnewline
\hline 
CoFeB, NiFe & 284--318 & \cite{neeraj2019experimental}\tabularnewline
\hline 
Py, Co & 0.83--3.1 & \cite{Li2015}\tabularnewline
\hline 
Co & 75--120 & \cite{unikandanunni2021inertial}\tabularnewline
\hline 
\multicolumn{3}{|c|}{Theory}\tabularnewline
\hline 
estimates & $\approx$1-100 & \cite{Ciornei2011, Bhattacharjee2012, Wegrowe2012}\tabularnewline
\hline 
bulk Fe, Co, Ni & $5.9-6.5\times10^{-3}$ & \cite{Thonig2017}\tabularnewline
\hline 
$3d$ and $4d$ impurities & $\approx$10--100 & \cite{Bouaziz2019}\tabularnewline
\hline 
\end{tabular}
\end{ruledtabular}

\end{table}


We note that non-harmonic time-dependent fields can also produce field-derivative torques, along with the ILLG spin dynamics. These spin torques are relativistic in nature and have been derived from Dirac theory \cite{Mondal2019PRB,Mondal2017Nutation}. 
\\
\\

\section{Inertial effects in ferromagnetic resonance}
\subsection{Ferromagnets}

For testing the accuracy of the model, it is necessary to connect the theoretical predictions based on the ILLG equation~\eqref{Eq5} to experimentally observable quantities. 
One of the possible methods is ferromagnetic resonance (FMR) where the linear response to a spatially homogeneous time-dependent external field is measured~\cite{Kittel1948}. A ferromagnet placed in a static external field $B_{\textrm{ext}}$ may be treated as a macrospin in FMR, which was investigated using numerical simulations of the ILLG equation in Ref.~\cite{Olive2012}. The magnetic susceptibility of the macrospin to a circularly polarized excitation 
of frequency $\omega$ is given by~\cite{cherkasskii2020nutation,Mondal2020nutation} 

\begin{equation} 
	{\chi (\omega)} = \dfrac{{\gamma M_{0}}}{{\gamma B_{\textrm{ext}} - \omega  - \eta {\omega ^2} + i\alpha \omega }}.
\end{equation}

\begin{figure}[tbh!]
    \centering
    \includegraphics[scale = 0.45]{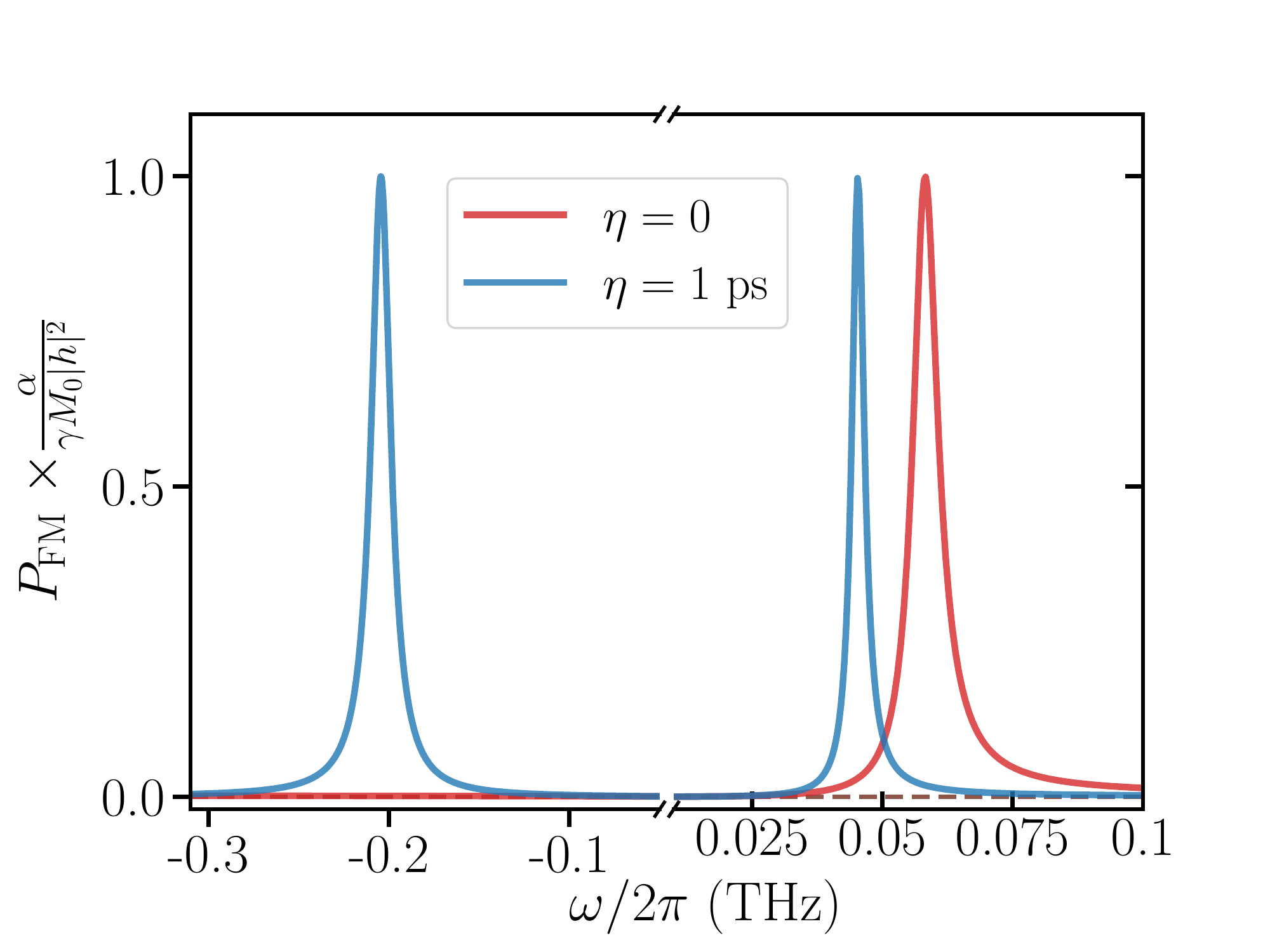}
    \caption{Ferromagnetic resonance for a single macrospin. The dissipated power compared between the inertia-free ($\eta=0$~ps) and inertial ($\eta=1$~ps) cases. The negative frequency of nutation indicates the opposite handedness of this motion compared to precession. The Hamiltonian is given by $\mathcal{H}=-M_{z}B_{\rm ext}-KM_{z}^{2}/M_{0}^{2}$. The calculation parameters are $ \gamma = 1.76\times 10^{11}$ T$^{-1}$s$^{-1}$, on-site anisotropy energy
$K = 10^{-23}$ {J},
$M_{ 0}  = 2 \mu_{\rm B}$, $ \alpha = 0.05$, and $ B_{\rm ext} = 1$ T. The data are taken from Ref. \cite{Mondal2020nutation}. 
    }
    \label{fig:suscep_FMR_nut}
\end{figure}

With the help of this susceptibility, one can calculate the dissipated power $P = \omega {\tt Im}[\chi(\omega)]$, shown in Fig.\,\ref{fig:suscep_FMR_nut}. The dissipated power shows peaks in the vicinity of the poles of the susceptibility,
\begin{equation} 
\begin{split}
{\omega _{\textrm{p}}} &= \frac{{-1 + \sqrt {1 + 4\eta \gamma B_{\textrm{ext}}} }}{{2\eta }} \\
&\approx \gamma B_{\textrm{ext}}\left(1-\eta\gamma B_{\textrm{ext}}\right)
\end{split}\label{eqprecFM}
\end{equation} 
and
\begin{equation}
\begin{split}
{\omega _{\textrm{n}}} &= \frac{{-1 - \sqrt {1 + 4\eta \gamma B_{\textrm{ext}}} }}{{2\eta }} \\
&\approx -\frac{1}{\eta }-\gamma B_{\textrm{ext}}\left(1-\eta\gamma B_{\textrm{ext}}\right)\,,
\end{split} \label{eqnutFM}
\end{equation}
where $\omega_{\rm p}$ and $\omega_{\rm n}$ denote the precession and nutation frequencies, respectively.
Approximate expressions for the frequencies were already derived in Ref.~\cite{Olive2015}, which reproduce the first term in the expansion.

{{The inertia causes a redshift of the precession frequency $\omega_{\textrm{p}}$ in Eq.~\eqref{eqprecFM}, as is visible in Fig.~\ref{fig:suscep_FMR_nut}. Unfortunately, this effect is not directly observable experimentally, since the inertia cannot be turned off in magnetic materials. Moreover, the resonance frequency may also be shifted by anisotropy effects discussed below, and the strength of the anisotropy terms would have to be also determined from the position of the FMR peak. As shown in Fig.~\ref{fig:FreqH}, the inertia also influences the dependence of the precession frequency on the external field. The effective gyromagnetic ratio $\gamma_{\textrm{eff}}=\partial\omega_{\textrm{p}}/\partial B_{\textrm{ext}}$ is decreased, and the frequency is no longer linear in the external field but also contains a term quadratic in $B_{\textrm{ext}}$ with a negative sign, which could represent an experimentally detectable signature.}} 
This direction was pursued in Ref.~\cite{Li2015}, where it was observed that the frequency is actually blueshifted at high fields, i.e., the coefficient of the $B_{\textrm{ext}}^{2}$ term is positive. Based on the energetical considerations fixing the sign of $\eta/\gamma>0$ discussed above, this seems to indicate that the redshift caused by the inertia is obscured by further effects not taken into account in Eq.~\eqref{eqprecFM}.

\begin{figure}
    \centering
    \includegraphics[width=\columnwidth]{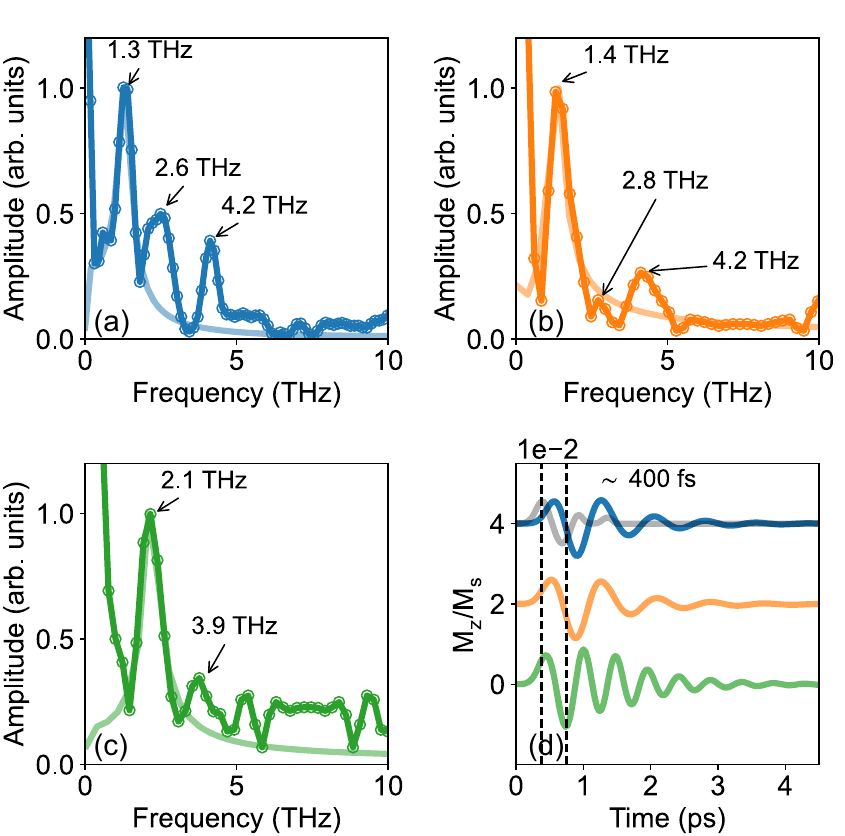}
    \caption{{{Experimental signatures of the inertial dynamics. Comparison between Fourier transforms of the probe signal in pump-probe measurements (solid line, open circles) and numerical simulations based on Eq.~\eqref{Eq5} for a single macrospin for (a) face-centered cubic (blue), (b) body-centered cubic (orange), and (c) hexagonal close-packed (green) cobalt thin films. The nutational resonance and its higher harmonics are visible in the experimental spectra. (d) Simulated response of the magnetization in the time domain. The semitransparent line shows the time integral of the pump pulse. Figure from Ref.~\cite{unikandanunni2021inertial}.}}
    }
    \label{fig:nut_exper}
\end{figure}

\begin{figure}[t]
    \centering
    \includegraphics[]{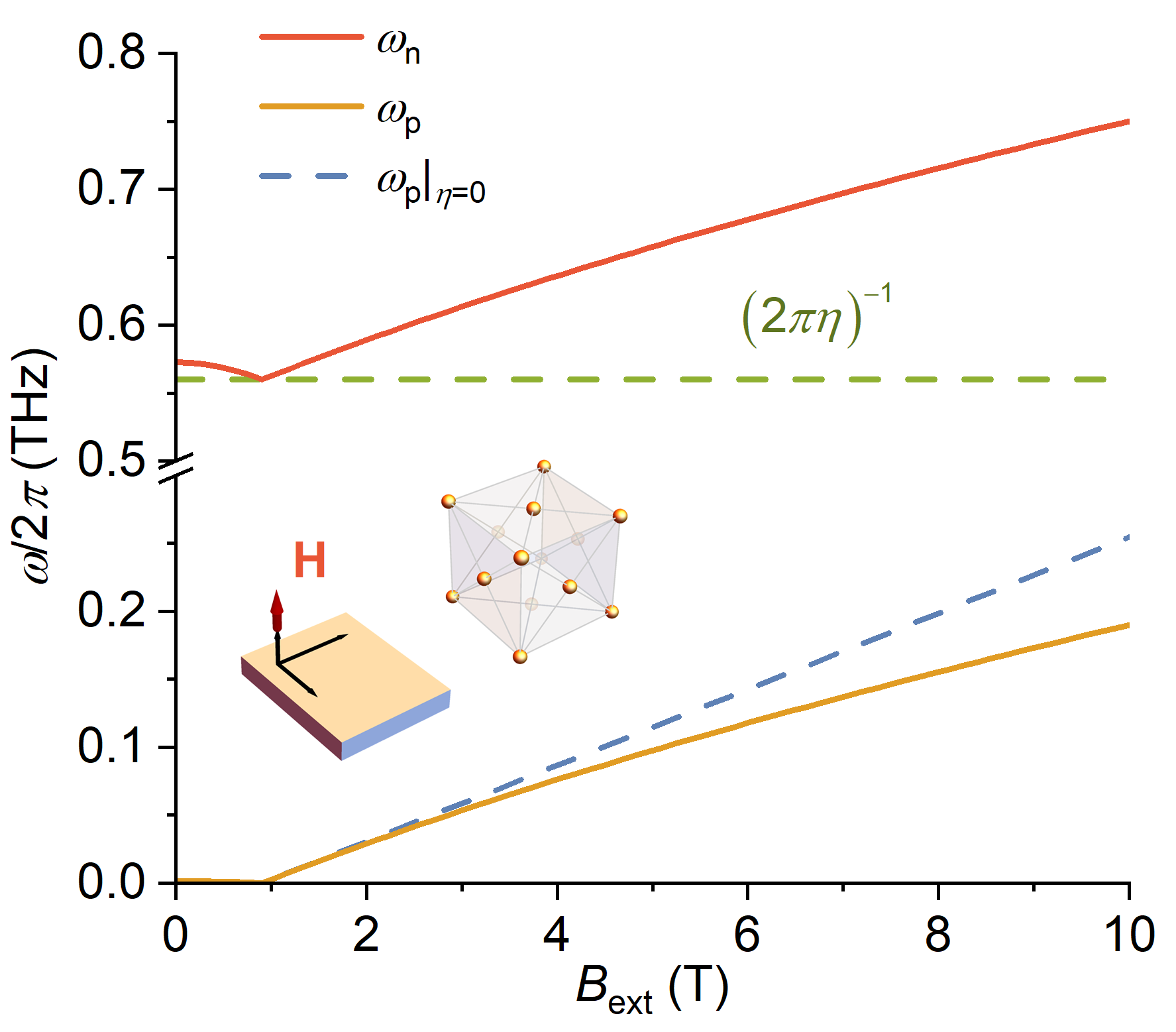}
    \caption{Frequency-field relation of precessional and nutational resonances. The explicit solution of Eq.\,\eqref{eq:SBC} for the precessional resonance (orange line) shows a redshift compared to the non-inertial Smit--Beljers case (dashed blue line), and the nutational resonance (red) demonstrates a blueshift compared to the zeroth-order approximation $1/\eta$. The calculation parameters for a thin film with cubic magnetocrystalline anisotropy are 
    ${\mu _0}{M_{\textrm{S}}} =0.9 \:\rm{T}$, $\alpha=0.0058$, $\eta = 284 \:\rm{fs}$, $K_{\rm{cub1}} =  4.9 \times 10^4 \: \rm{ J} \cdot \rm{m}^{-3}$. The data are taken from\,Ref.\,\cite{Cherkasskii2022Anisotropy}}
    \label{fig:FreqH}
\end{figure}

More promising for the observation is the emergence of a second nutational resonance peak $\omega_{\textrm{n}}$ in Fig.~\ref{fig:suscep_FMR_nut}. The negative frequency denotes the opposite handedness of this excitation compared to the counterclockwise precession~\cite{Kikuchi,Mondal2020nutation}. Since the LLG equation is not capable of explaining the emergence of magnetic excitations in ferromagnets at such high frequencies, a resonant excitation at the nutation frequency provides a distinct signature of inertial dynamics. Such high-frequency resonances were  recently detected via time-resolved magneto-optical pump-probe techniques in Refs.~\cite{neeraj2019experimental,unikandanunni2021inertial}. { {Typical experimental spectra from Ref.~\cite{unikandanunni2021inertial} are shown in Fig.\,~\ref{fig:nut_exper}. The presence of higher harmonics in the experimental signal not predicted by linear-response theory might represent an indication of non-linear processes. 
}}

{{
A further possibility for the experimental investigation of the nutation resonances is based on spin pumping. In this case, a spin current is injected from an externally excited ferromagnet into an adjacent non-magnetic metal. This spin current is predicted to change sign as the frequency is changed from the precession to the nutation resonance~\cite{Mondal2021PRBSpinCurrent}.}}

The inertial dynamics of a general anisotropic macrospin in the linear-response formalism was investigated in Refs.~\cite{Cherkasskii2022Anisotropy, titov2022ferromagnetic}. 
After expressing the free-energy density $F$ in polar coordinates $\vartheta$ and $\varphi$, the excitation frequencies are found from the solution of the fourth-order secular equation

    \begin{equation} \label{eq:SBC}
    \begin{split}
		 & \left[ \frac{\omega ^2}{\gamma^2} - \frac{{\left( {1 + {\alpha ^2}} \right)}}{{M_{\textrm{S}}^2{{\sin }^2}{\vartheta}}}\left( {{\partial _{\vartheta \vartheta }}F{\partial _{\phi \phi }}F - {{\left( {{\partial _{\vartheta \phi }}F} \right)}^2}} \right) \right] \\
		 & - {\eta ^2 \omega ^2} \left[ \frac{{\omega ^2}}{ \gamma ^2} - \frac{1}{\eta \gamma M_{\textrm{S}}}\left( {{\partial _{\vartheta \vartheta }}F + \frac{{{\partial _{\varphi \varphi }}F}}{{{{\sin }^2}{\vartheta}}}} \right) \right]\\
		 & - i\omega \frac{{\alpha}}{\gamma M_{\textrm{S}}}\left( {{\partial _{\vartheta \vartheta }}F + \frac{{{\partial _{\varphi \varphi }}F}}{{{{\sin }^2}{\vartheta}}}} \right) = 0,
	\end{split}
    \end{equation}
The first 
line corresponds to the Smit--Beljers equation. The second 
line includes the inertia. The third 
line induces the frequency-domain linewidth of the FMR. Note that the solutions of Eq.~\eqref{eq:SBC} can be grouped into pairs of $\omega$ and $-\omega^{*}$ due to a particle-hole constraint, and the two frequency pairs describe precessional and nutational excitations. 

For instance, if the magnetic field $B_{\textrm{ext}}$ is applied out-of-plane with respect to the surface of a film demonstrating cubic magnetocrystalline anisotropy $K_{\textrm{cub1}}$, 
the free-energy density is given by

\begin{equation}
\begin{split}
F= & -B_{{\rm ext}}M_{z}+\dfrac{1}{2}\mu_{0}M_{z}^{2}\\
 & +\frac{K_{{\rm cub1}}}{M_{\textrm{S}}^{4}} \left(M_{x}^{2}M_{y}^{2}+M_{y}^{2}M_{z}^{2}+M_{x}^{2}M_{z}^{2}\right),
\end{split}
\end{equation}

allowing to find the approximate solution of Eq.~\eqref{eq:SBC}:
\begin{equation} \label{eq:w_approx_prec_b_cub_H_OOP}
    \begin{split}
    & {\omega_{\rm{p}}}^2 \approx  
    \gamma^2\left(1+ \alpha ^2 \right) \left(-\mu_0 M_{\textrm{S}} + B_{\textrm{ext}}  + \dfrac{2 K_{\rm{cub1}}}{M_{\textrm{S}}} \right)^2 \\
    & \times \left[1-\eta\gamma\left(-2 \mu_0 M_{\textrm{S}} +2 B_{\textrm{ext}}  + \dfrac{4 K_{\rm{cub1}}}{M_{\textrm{S}}}\right)\right],
    \end{split}
    \end{equation}

\begin{equation} \label{eq:w_approx_b_nut_cub_H_OOP}
	{\omega}_{\rm{n}} \approx \frac{1}{\eta } + \gamma\left( { - {\mu _0}{M_{\textrm{S}}} + B_{\textrm{ext}} + \frac{{2{K_{{\rm{cub1}}}}}}{{{M_{\textrm{S}}}}}} \right).
\end{equation}

The numerical solutions of Eq.\,(\ref{eq:SBC}) are plotted in Fig.\,\ref{fig:FreqH}. In agreement with Eqs.~\eqref{eqprecFM} and \eqref{eqnutFM}, this approximation shows a redshift for the precessional resonance compared to the non-inertial Smit--Beljers case, and a blueshift of the nutational resonance compared to the zeroth-order approximation $1/\eta$. 

In the undamped limit of the non-linear ILLG equation, analytic solutions for the magnetization of a macrospin with uniaxial magnetocrystalline anisotropy parallel to the external field direction were obtained in terms of the Jacobi elliptic functions and elliptic integrals in Ref.~\cite{Titov2021Deterministic}. In this work, the nutation frequency was determined in terms of the inverse period of the Jacobi elliptic function. In addition, the equilibrium correlation functions of the magnetization at short times were investigated in Ref.~\cite{TitovCorrelation2023}.

For the sake of completeness, it should be mentioned that high resonance frequencies have also been predicted based on the conventional LLG equation due to surface anisotropy effects in ferromagnetic nanoparticles~\cite{Bastardis2018}. These must be distinguished from the resonances caused by the inertial motion of a homogeneous magnetization discussed here.

\subsection{Antiferromagnets and ferrimagnets}

Unlike ferromagnets, antiferromagnets and ferrimagnets consist of 
multiple 
magnetic sublattices 
pointing along different directions. 
In two-sublattice systems, the magnetic susceptibility shows two resonances with opposite handedness, similarly to the precessional and nutational resonances in ferromagnets. Furthermore, the antiferromagnetic exchange coupling typically shifts both resonances in antiferromagnets 
and one of the resonances in ferrimagnets to the THz regime~\cite{Kittel1951,Takeo1951,Keffer1952}. The other ferrimagnetic resonance usually resides in the GHz range, because ferrimagnets have a net magnetic moment similarly to ferromagnets. 

It was investigated in Ref.~\cite{Mondal2020nutation} how these resonances are influenced by the inertia. Two-sublattice systems may be treated as two interacting macrospins $\boldsymbol{M}_{A}$ and $\boldsymbol{M}_{B}$ described by the Hamiltonian
\begin{align}
\mathcal{H}=&-\left(M_{A,z}+M_{B,z}\right)B_{\textrm{ext}}-\frac{K_{A}}{M_{\textrm{0},A}^{2}}M_{A,z}^{2}-\frac{K_{B}}{M_{\textrm{0},B}^{2}}M_{B,z}^{2} \nonumber
\\
&+\frac{J}{M_{\textrm{0},A}M_{\textrm{0},B}}\boldsymbol{M}_{A}\cdot\boldsymbol{M}_{B}\, ,\label{eqAFMHam}
\end{align}
where $K_{A/B}$ are the uniaxial anisotropy coefficients, $M_{0,A}$ and $M_{0,B}$ are the sizes of the magnetic moments and $J$ is the exchange interaction. The linear response is calculated around the state where $\boldsymbol{M}_{A}$ and $\boldsymbol{M}_{B}$ are parallel and antiparallel to the external field, respectively. The resonance frequencies can be identified as the poles of the susceptibility tensor. Although Eq.~\eqref{eqAFMHam} possesses a cylindrical symmetry simplifying the calculations, this requires solving a fourth-order algebraic equation due to the two sublattices. 

\begin{figure}[tbh!]
    \centering
    \includegraphics[scale = 0.5]{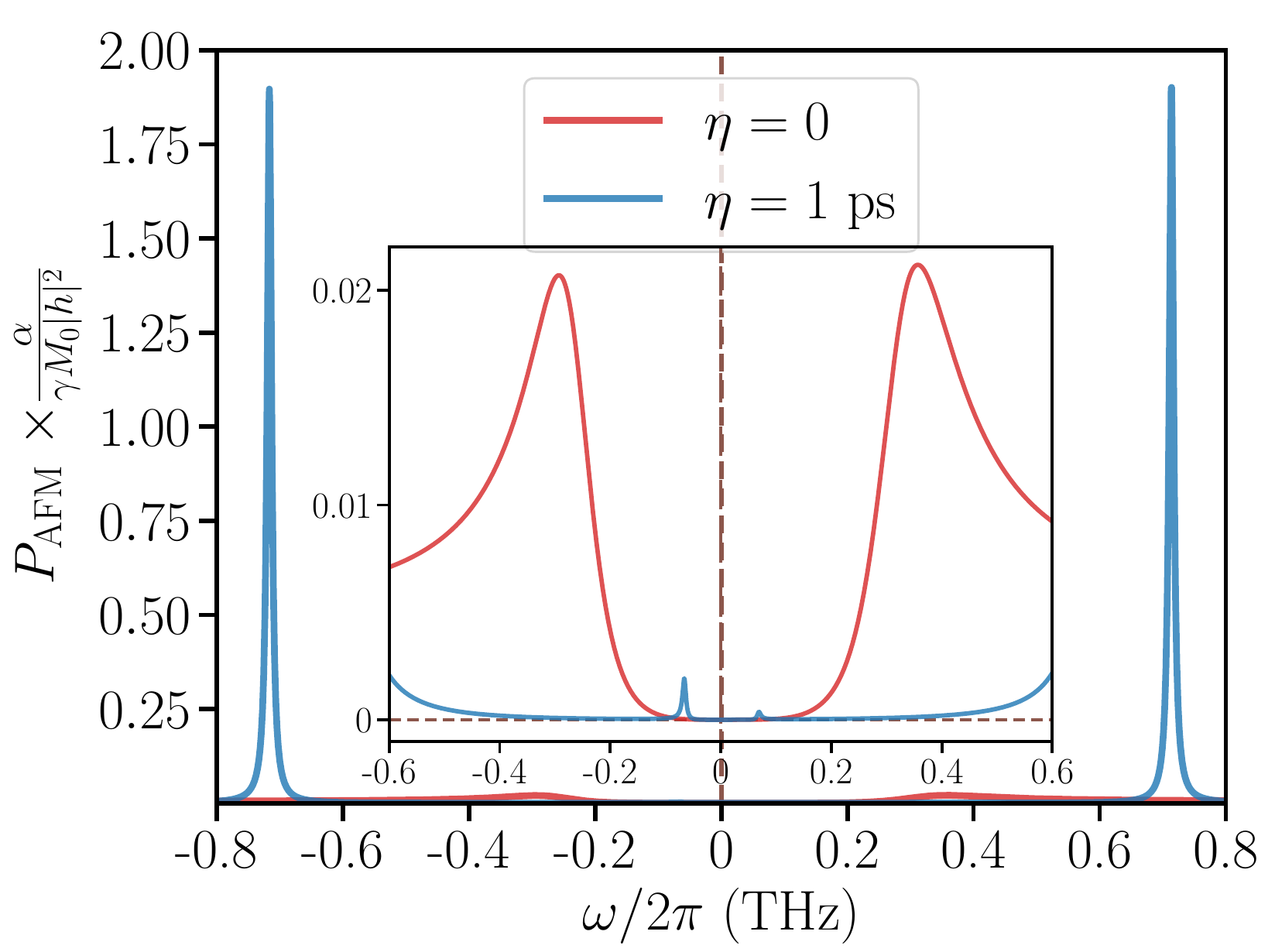}
    \caption{The precessional and nutational resonance frequencies for antiferromagnets. The dissipated power is compared between the inertia-free ($\eta=0$~ps) and inertial ($\eta=1$~ps) cases. The calculation parameters are $\gamma_A = \gamma_B   = \gamma = 1.76\times 10^{11}$ T$^{-1}$s$^{-1}$, $J  = 10^{-21}$ {J},
$K_A = K_B = K = 10^{-23}$ {J},
$M_{0,A}  = M_{0,B} = 2 \mu_{\rm B}$, $\alpha_{A}  = \alpha_{B} = \alpha = 0.05$, and $ B_{\rm ext} = 1$ T. The data are taken from Ref. \cite{Mondal2020nutation}.}
    \label{fig:fig4}
\end{figure}

In the antiferromagnetic limit with $\eta_{A}=\eta_{B}=\eta$, 
$\gamma_{A}=\gamma_{B}=\gamma$, $M_{\textrm{0},A}=M_{\textrm{0},B}=M_{\textrm{0}}$ and $K_{A}=K_{B}=K$, the undamped excitation frequencies may be approximated as~\cite{Mondal2020nutation}
\begin{align}
\omega_{\textrm{p}\pm}=&\pm\frac{\gamma}{M_{\textrm{0}}}\frac{\sqrt{4KJ}}{\sqrt{1+2\eta\gamma J/M_{\textrm{0}}}}+\frac{\gamma B_{\textrm{ext}}}{1+2\eta\gamma J/M_{\textrm{0}}}\, ,\label{eqprecAFM}
\\
\omega_{\textrm{n}\pm}=&\pm\frac{1}{\eta}\sqrt{1+2\eta\gamma J/M_{\textrm{0}}}-\frac{\gamma B_{\textrm{ext}}}{1+2\eta\gamma J/M_{\textrm{0}}}\, ,\label{eqnutAFM}
\end{align}
for $K,M_{\textrm{0}}B_{\textrm{ext}}\ll J$. These resonances are observable as peaks in the dissipated power in Fig.~\ref{fig:fig4}. Similarly to ferromagnets, the number of peaks is doubled and the precessional resonance frequency is reduced with the introduction of magnetic inertia. Note that the redshift of the precessional resonance frequency is determined by the dimensionless parameter $\gamma\eta J/M_{0}$ in antiferromagnets and $\gamma\eta B_{\textrm{ext}}$ in ferromagnets, therefore it is exchange enhanced in the former. As mentioned in the ferromagnetic case, this shift itself is not detectable experimentally since it is not possible to distinguish the influence of inertia from, e.g., a different value of the anisotropy. Furthermore, antiferromagnetic precessional resonance peaks have a much lower intensity as can be seen from the comparison between Fig.~\ref{fig:suscep_FMR_nut} and Fig.~\ref{fig:fig4}. However, the nutational peaks have a much higher intensity and a sharp lineshape, since the exchange enhancement of the effective damping parameter only affects the precessional peaks~\cite{Mondal2020nutation}. Although so far there are no experimental investigations of the magnetic inertia in antiferromagnets reported in the literature, these properties indicate that observing the nutational resonances would be possible in them just as 
in ferromagnets. Furthermore, since the precessional resonances also have higher frequencies, the same THz methods could be used for the detection of precessional and nutational resonances, while the GHz precessional frequencies in ferromagnets are typically measured using a different approach. 
\begin{figure}[tbh!]
    \centering
    \includegraphics[scale = 0.25]{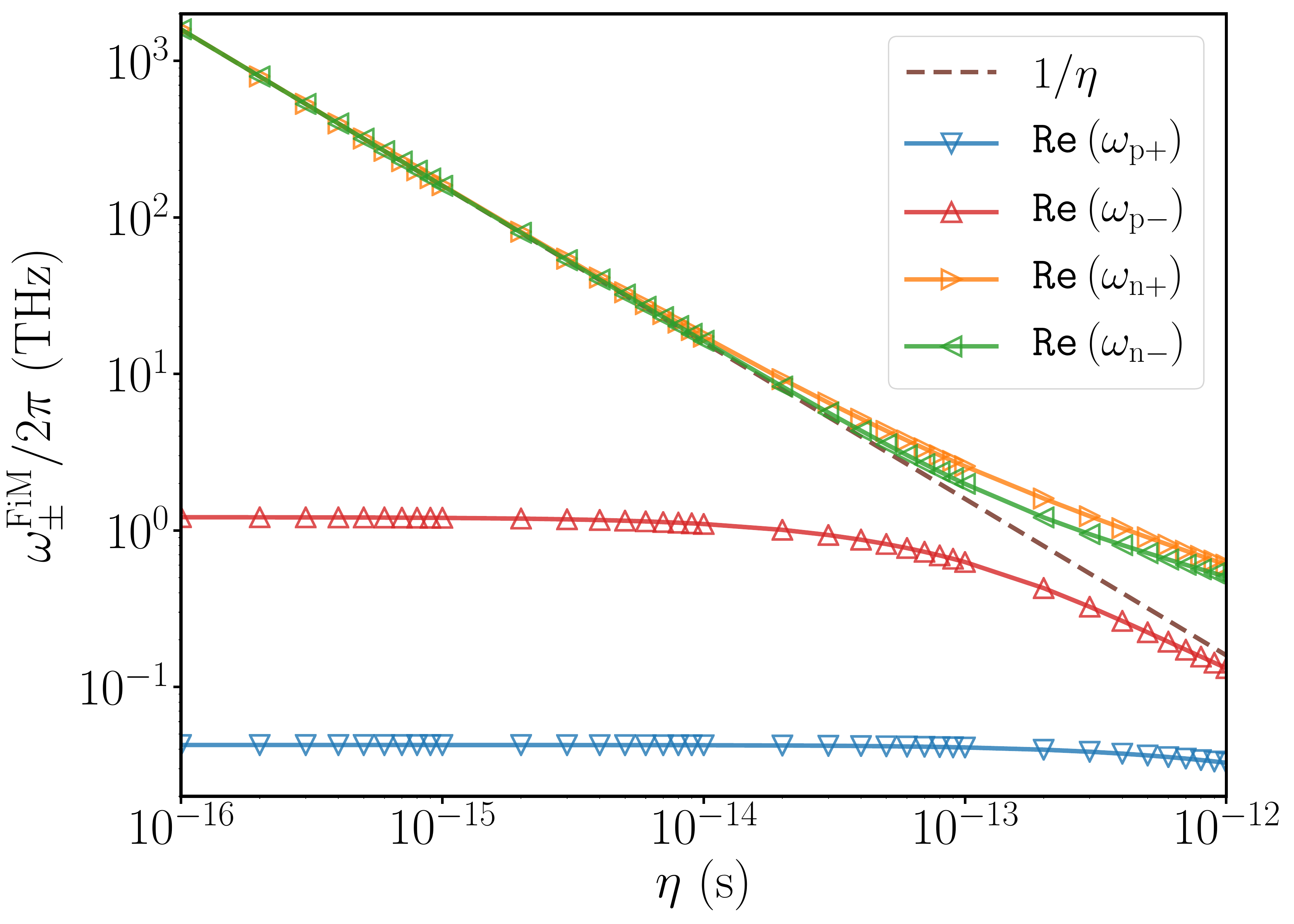}
    \caption{The precessional and nutational resonance frequencies for ferrimagnets as a function of the inertial parameter $\eta$. The calculation parameters are $M_{0,A} = 2\mu_B, M_{0,B} = 5M_{0,A} = 10\mu_B, \gamma_A = \gamma_B  = 1.76\times 10^{11}$ T$^{-1}$s$^{-1}$, $J  = 10^{-21}$ {J},
$K_A = K_B  = 10^{-23}$ {J},
$\alpha_{A}  = \alpha_{B}  = 0.05$, and $ B_{\rm ext} = 1$ T. The figure is taken from Ref. \cite{Mondal2020nutation}.}
    \label{fig:fig5}
\end{figure}

The numerically calculated resonance frequencies in a ferrimagnet are displayed in Fig.~\ref{fig:fig5}. The precessional frequency $\omega_{\textrm{p}+}$ only starts to be influenced by the nutational frequency $\omega_{\textrm{n}-}$ for large values of $\eta$, similarly to the ferromagnetic case. In contrast, the strong interaction between the $\omega_{\textrm{p}-}$ and $\omega_{\textrm{n}+}$ frequencies resembles the antiferromagnetic case. Therefore, the same considerations as above concerning the possible experimental detection of the nutational resonance apply here.

\section{Nutational spin waves}
\begin{figure*}[tbh!]
    \centering
\includegraphics[width=1\textwidth]{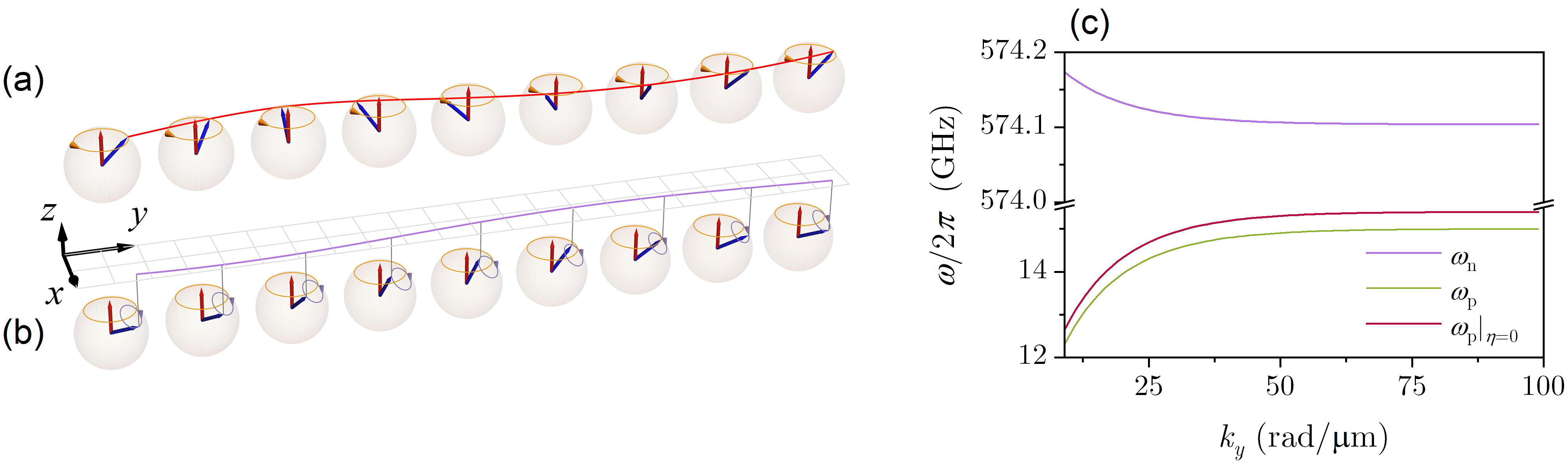}
    \caption{(a) Precessional spin wave without inertia. 
    The blue arrows indicate the motion of the magnetic moments $\boldsymbol{M}_{i}$  in a ferromagnet. (b) Nutational 
    spin wave 
    with a frequency considerably higher than in (a) plotted with small blue circles on top of the ``frozen'' precessional motion. Panels (a) and (b) from Ref. \cite{Cherkasskii2021}. (c) The dispersion branches of nutational surface spin waves in the THz 
    range 
    $\omega_{\textrm{n}}$, the precessional Damon-Eshbach mode without inertia 
    $\omega_{\textrm{p}}|_{\eta=0}$, and the precessional Damon-Eshbach mode shifted by inertia 
    $\omega_{\textrm{p}}$. 
    The calculation parameters for the thin film are: 
    ${\mu _0}{M_{\textrm{S}}} =0.9 \:\rm{T}$,  $B_{\textrm{ext}} =0.1 \:\rm{T}$, $\alpha=0.0058$, $\eta = 284 \:\rm{fs}$, film thickness $L =  40\: \rm{nm}$.}
    \label{fig:NutWaves}
\end{figure*}
Due to the interaction between the magnetic moments, the linearized ILLG equation also possesses propagating solutions known as spin waves, illustrated in Fig.\,\ref{fig:NutWaves}(a). 
Taking inertia into account, nutational spin waves also appear alongside the conventional  precessional spin waves. 
Since nutational spin waves have THz frequencies compared to the typically GHz frequencies of the precessional spin-wave modes in ferromagnets, they can be imagined as a small deviation on top of a ``frozen'' precessional motion, shown in Fig.\,\ref{fig:NutWaves}(b). Conventionally the spin-wave dispersion relation is separated into two regimes: at long wave vectors comparable to the sample sizes magnetostatic effects dominate, while at shorter wavelengths the short-ranged exchange interactions play the most important role.

The magnetostatic nutational waves were studied in in-plane magnetized ferromagnetic thin films in Ref.~\cite{Cherkasskii2021}. It was found that for the spin waves propagating perpendicular to the applied magnetic field (Damon--Eshbach configuration), inertial effects on magnons are twofold: the frequency of precessional waves is reduced and 
nutational surface spin waves emerge, as shown in Fig.~\ref{fig:NutWaves}(c). Notably, nutational spin waves propagate with a group velocity opposite to their wave vector, which is only observed for precessional spin waves with wave vectors parallel to the magnetic field (backward volume modes). 
The interaction of spin waves described by the ILLG equation with electromagnetic waves in ferromagnets was investigated 
in Ref.~\cite{Titov2022NutationWaves}. The interaction leads to the hybridization between magnons and photons and the opening of avoided crossings in the spectrum. Since the typical nutational spin-wave frequencies are in the range of $\eta^{-1}\sim 10^{13}-10^{15}$~s$^{-1}$, the wave vectors of electromagnetic waves hybridizing with these modes are around $k\sim 10^{5}-10^{7}$~m$^{-1}$, falling into the magnetostatic regime.

\begin{figure}[t]
    \centering
    \includegraphics[]{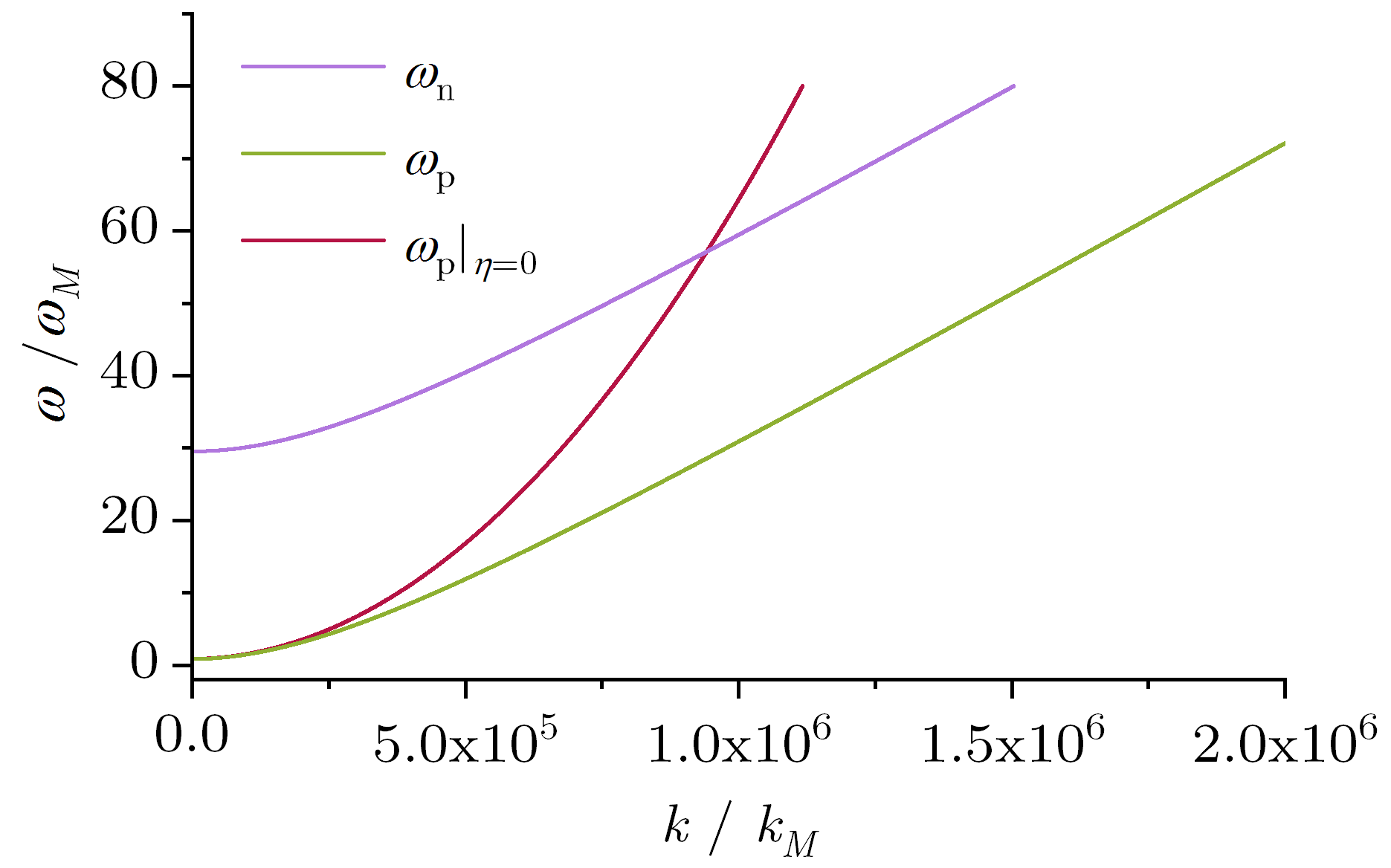}
    \caption{Dispersion relation of exchange spin waves without and with inertia in ferromagnets: nutational spin waves 
    $\omega_{\textrm{n}}$, inertial precessional spin waves 
    $\omega_{\textrm{p}}$, and non-inertial precessional spin waves 
    $\omega_{\textrm{p}}|_{\eta=0}$. Here $\omega_M = \gamma  \mu_0 M_{\textrm{S}}$, $k_M = \sqrt{\varepsilon_r} \omega_M / c$, where $c$ is the speed of light and $\varepsilon_r$ is the relative permittivity of ferromagnets. The calculation parameters are the following: 
    ${\mu _0}{M_{\textrm{S}}} =0.2 \:\rm{T}$,  $B_{\textrm{ext}} =0.1 \:\rm{T}$, $\eta = 1 \:\rm{ps}$, and $\varepsilon_r =  15.5$.}
    \label{fig:NutWaves_Titov}
\end{figure}

Nutational exchange spin waves were discussed in Refs.~\cite{Kikuchi,Makhfudz2020,Titov2022NutationWaves,Lomonosov2021,Mondal2022}. For a nearest-neighbour ferromagnetic exchange interaction $J$, the dispersion relation may be approximated in the long-wavelength regime 
as
\begin{align}
\omega_{\textrm{p},\boldsymbol{k}}\approx &\frac{zJa^{2}}{2}\boldsymbol{k}^{2}\left(1-\eta \frac{zJa^{2}}{2}\boldsymbol{k}^{2}\right)\, ,\label{eq:ferroprecwave}
\\
\omega_{\textrm{n},\boldsymbol{k}}\approx &-\frac{1}{\eta}-\frac{zJa^{2}}{2}\boldsymbol{k}^{2}\left(1-\eta \frac{zJa^{2}}{2}\boldsymbol{k}^{2}\right)\, \label{eq:ferronutwave}
\end{align}
for precessional and nutational spin waves, respectively. Here, $z$ is the number of nearest neighbours and $a$ is the distance between the corresponding sites. The negative sign of the nutational frequency indicates an opposite handedness compared to the precessional waves~\cite{Kikuchi}, as already mentioned for the $\boldsymbol{k}=\boldsymbol{0}$ FMR mode in Eqs.~\eqref{eqprecFM} and \eqref{eqnutFM}. If the spin-wave dispersion becomes non-reciprocal, for example due to the presence of the Dzyaloshinskii--Moriya interaction, the opposite handedness gives rise to a minimum in the dispersion relation for opposite wave vectors in the two branches~\cite{Mondal2022}. The precessional and nutational branches differ by a constant shift $\eta^{-1}$, and the frequencies of the precessional modes are decreased due to the inertia, as illustrated in Fig.~\ref{fig:NutWaves_Titov}. 

Although the dispersion relation of the precessional spin waves is modified by inertia, similar frequency shifts may also be explained within the LLG equation by choosing a different saturation magnetization, magnetocrystalline anisotropy term or taking exchange interactions with further neighbours into account. Unless these parameters are known from independent measurements of static properties which are not expected to be affected by the inertia, such as the temperature dependence of the magnetization or the critical temperature, a measurement of the precessional branch only is unlikely to result in a convincing indication for the prevalence of inertial phenomena. The same argument holds for the group velocity, the gyromagnetic ratio or the effective damping parameter which are also influenced by the inertia~\cite{Mondal2022,Lomonosov2021,Makhfudz2020}, as mentioned above in the case of the ferromagnetic resonance. Experimental results on nutational spin waves are not available at the moment, but they could provide sufficient evidence for the theoretical predictions based on the ILLG equation. Of particular interest would be the investigation of nutational spin-wave modes with a group velocity opposite to their wave vector, as discussed in the Damon--Eshbach configuration for ferromagnets above and for exchange spin waves in antiferromagnets in Ref.~\cite{Mondal2022}.

\section{Magnetization switching in the inertial regime}

While linear-response and linear spin-wave theories describe the time evolution close to the equilibrium state, the switching between different equilibrium states is a non-linear effect that is also influenced by the inertial dynamics. The ILLG equation was solved numerically for a single uniaxial macrospin under the influence of a magnetic field pulse with zero frequency perpendicular to the easy-axis direction in Ref.~\cite{neeraj2021magnetization}. The switching time was found to be lower in a wide range of pulse durations in the inertial case than for the non-inertial LLG equation. However, as emphasized before such a quantitative effect may be difficult to probe experimentally where inertial and non-inertial dynamics cannot be compared directly. 

\begin{figure}[tbh!]
    \centering
    \includegraphics[width=\columnwidth]{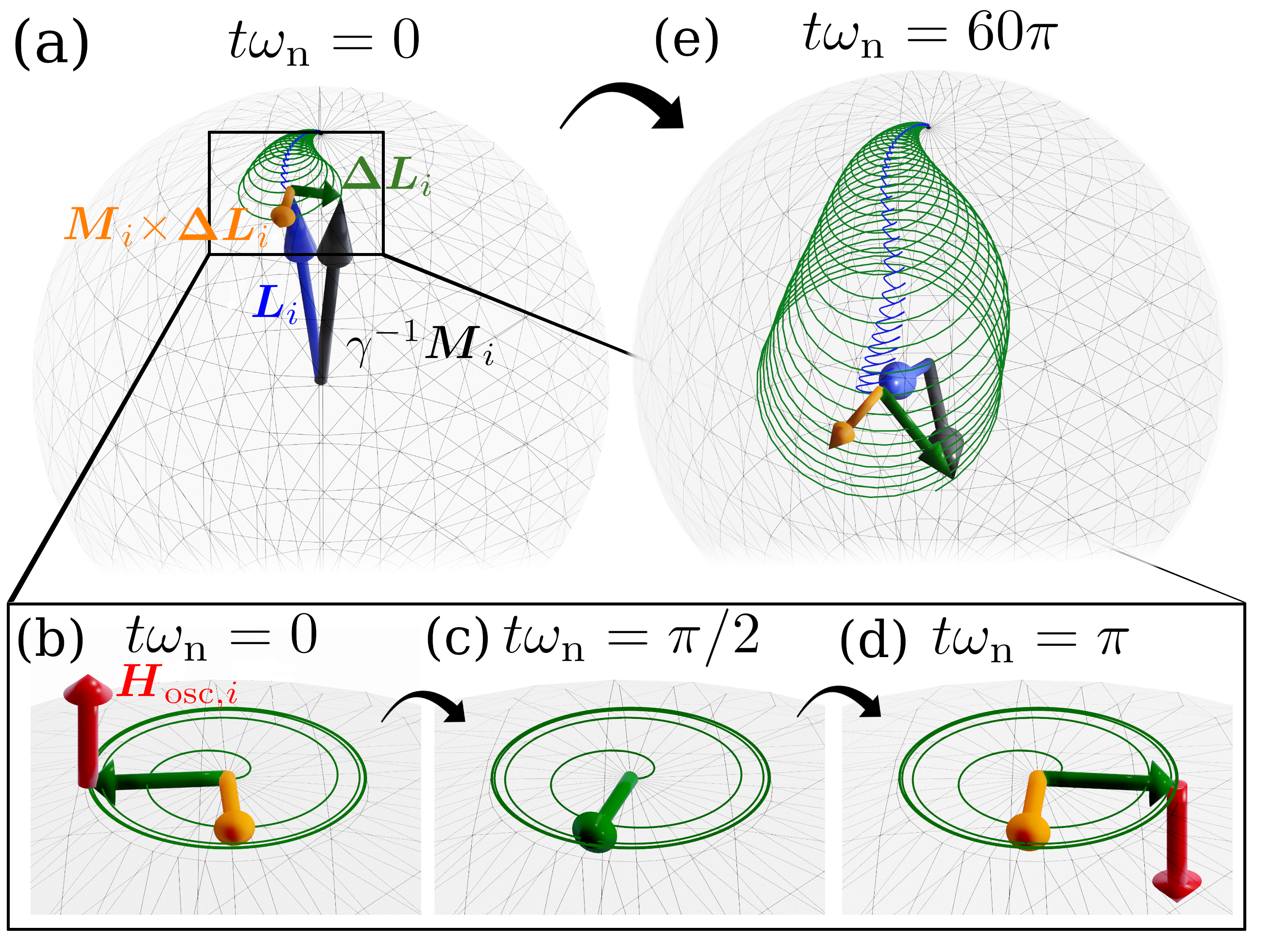}
    \caption{Illustration of magnetization switching in the inertial regime. (a) The angular momentum $\boldsymbol{L}_{i}$ and the magnetic moment $\gamma^{-1}\boldsymbol{M}_{i}$ differ by the nutation vector $\Delta\boldsymbol{L}_{i}$ due to the inertial dynamics. The nutation is excited by the oscillating external magnetic field $\mu_{0}\boldsymbol{H}_{\textrm{osc}}$ resonantly at the nutation frequency $\omega_{\textrm{n}}$. (b)-(d) The oscillating field exerts a torque $-\gamma\Delta\boldsymbol{L}_{i}\times\mu_{0}\boldsymbol{H}_{\textrm{osc}}$ on the angular momentum. Since the nutation vector follows the excitation field with a finite phase shift, the average of the torque over a period of the excitation is finite. (e) Over several nutation periods, the torque causes a switching of the angular momentum and the magnetic moment. Figure from Ref.~\cite{Winter2022}.}
    \label{fig:fig8}
\end{figure}

In Ref.~\cite{Winter2022}, it was found by combining analytical calculations and numerical simulations that a resonant excitation of the nutation amplitude gives rise to a torque that is capable of switching the magnetic moment, which effect is unparalleled in the LLG equation. This phenomenon is illustrated in Fig.~\ref{fig:fig8}. Since the switching velocity was found to be proportional to the square of the nutation amplitude which scales with the amplitude of the ac excitation field itself, the velocity increases quadratically with the field amplitude instead of linearly in the case of switching based on the Larmor precession. This enables lower switching times compared to precessional switching for intermediate field strengths. Furthermore, it was demonstrated that both $90^{\circ}$ and $180^{\circ}$ switching may be achieved for a single macrospin depending on the linear or circular polarization of the excitation field, while in antiferromagnets switching to states with the magnetic moments either perpendicular to or in the plane of the excitation field may be realized depending on the excitation frequency. 

Based on the solution of the Fokker--Planck equation for the inertial Landau--Lifshitz--Gilbert--Bloch equation, it was argued in Ref.~\cite{Makhfudz2022} that inertial effects together with thermal excitations could explain puzzling effects observed in all-optical magnetization switching, including its dependence on the polarization of the laser pulse.


\section{Conclusion}

We reviewed how the inclusion of an inertial term in the quasiclassical Landau--Lifshitz--Gilbert equation influences the resonance frequencies and the spin-wave modes in the linear-response regime, as well as the switching times in magnetic nanoparticles at ultrafast time scales. Apart from quantitative changes compared to the Landau--Lifshitz--Gilbert dynamics, the inertia gives rise to qualitatively new phenomena such as nutational spin waves and resonances, the excitation of which opens up faster paths for the reversal of the magnetic moments. Precisely these new phenomena provide the most promising way to detect signatures of the inertia, since the quantitative changes may also be interpreted using a different choice of parameters within the LLG dynamics, and the values of these parameters are not known a priori.

Experimental observations of these effects have been restricted to nutational resonances in ferromagnets so far~\cite{neeraj2019experimental,unikandanunni2021inertial}, which have considerably higher frequencies compared to the typical precessional modes. Optical methods appear to be particularly suitable for obtaining further experimental evidence on inertial effects, because the frequency of the electromagnetic waves used in these methods falls into the range where nutational spin waves are expected to emerge, and the possible control over their polarization may enable different switching paths. Antiferromagnets may be particularly appealing for this purpose, since precessional and nutational spin waves in them are located in the same frequency range, possibly enabling their simultaneous observation using the same technique. The requirement for adjusting the frequency represents a challenge, in particular because estimates for the nutational resonance frequency in the literature differ by two orders of magnitude for similar materials. Further first-principles calculations of the inertial relaxation time may provide guidance on the choice of materials for the experiments. Theoretical calculations based on microscopic models of the interactions of the magnetic degrees of freedom with electrons, phonons or photons could provide important comparisons with the quasiclassical description discussed here. A joint effort from the experimental and theoretical sides is expected to provide valuable insight into the limits of applicability of inertial spin dynamics at femtosecond time scales and beyond.



\section*{Acknowledgments}
We are grateful to Anna Semisalova, Igor Barsukov, Jean-Eric Wegrowe and Sebastian T. B. Goennenwein for fruitful discussions. Financial support by the faculty research scheme at IIT (ISM) Dhanbad, India under Project No. FRS(196)/2023-2024/PHYSICS, by the National Research, Development, and Innovation Office (NRDI) of Hungary under Project Nos. K131938 and FK142601, by the Young Scholar Fund at the University of Konstanz, by the Ministry of Culture and Innovation; National Research, Development and Innovation Office within the Quantum Information National Laboratory of Hungary (Grant No. 2022-2.1.1-NL-2022-00004), by the Swedish Research Council (VR), and by the K. and A. Wallenberg Foundation (Grant No. 2022.0079) is gratefully acknowledged.



\providecommand{\noopsort}[1]{}\providecommand{\singleletter}[1]{#1}%

\end{document}